\documentclass[10pt,conference]{IEEEtran}
\IEEEoverridecommandlockouts
\usepackage{cite}
\usepackage{amsmath,amssymb,amsfonts}
\usepackage{algorithm} 
\usepackage{algpseudocode}
\usepackage{graphicx}
\usepackage{textcomp}
\usepackage{multirow}
\usepackage[table,xcdraw]{xcolor}
\usepackage{caption}
\usepackage{array}
\usepackage{booktabs}
\usepackage{float}
\usepackage{balance}
\usepackage{enumitem}
\usepackage{listings}
\usepackage{subcaption}
\usepackage{booktabs}          
\usepackage{colortbl}     
\usepackage{xcolor}          
\usepackage{amssymb}  
\newcommand{\cmark}{\checkmark}  
\newcommand{\xmark}{\texttimes}  
\usepackage{tikz}
\usetikzlibrary{arrows,automata}
\usepackage{textcomp}
\usepackage{multirow}
\usepackage{pgfplots}
\pgfplotsset{compat=1.17} 
\usepgfplotslibrary{groupplots}
\usepackage{adjustbox}
\usepackage{empheq}

\def\BibTeX{{\rm B\kern-.05em{\sc i\kern-.025em b}\kern-.08em
    T\kern-.1667em\lower.7ex\hbox{E}\kern-.125emX}}

\lstdefinestyle{mystyle}{
    backgroundcolor=\color{backcolour},   
    commentstyle=\color{codegreen},
    keywordstyle=\color{magenta},
    numberstyle=\tiny\color{codegray},
    stringstyle=\color{codepurple},
    basicstyle=\ttfamily\footnotesize,
    breakatwhitespace=false,         
    breaklines=true,                 
    captionpos=b,                    
    keepspaces=true,                 
    numbers=left,                    
    numbersep=5pt,                  
    showspaces=false,                
    showstringspaces=false,
    showtabs=false,                  
    tabsize=2
}

\begin{document}

\title{Tetris: An SLA-aware Application Placement Strategy in the Edge-Cloud Continuum}

\author{
    \IEEEauthorblockA{Lucas Almeida\IEEEauthorrefmark{1} ,  Maycon Peixoto\IEEEauthorrefmark{1}} \\\
    \IEEEauthorblockA{\IEEEauthorrefmark{1}Institute of Computing, Federal University of Bahia, Brazil \{lucasmascarenha, maycon.leone\}@ufba.br}
}


\maketitle

\begin{abstract}
An Edge-Cloud Continuum integrates edge and cloud resources to provide a flexible and scalable infrastructure. This paradigm can minimize latency by processing data closer to the source at the edge while leveraging the vast computational power of the cloud for more intensive tasks. In this context, module application placement requires strategic allocation plans that align user demands with infrastructure constraints, aiming for efficient resource use. Therefore, we propose Tetris, an application placement strategy that utilizes a heuristic algorithm to distribute computational services across edge and cloud resources efficiently. Tetris prioritizes services based on SLA urgencies and resource efficiency to avoid system overloading. Our results demonstrate that Tetris reduces SLA violations by approximately 76\% compared to the baseline method, which serves as a reference point for benchmarking performance in this scenario. Therefore, Tetris offers an effective placement approach for managing latency-sensitive applications in Edge-Cloud Continuum environments, enhancing Quality of Service (QoS) for users.
\end{abstract}


\section{Introduction}
\label{introducao}

Edge computing is a paradigm designed to address the increasing computational demands and latency constraints of modern applications. By processing data closer to the source, edge computing can significantly reduce latency, enhance privacy, and improve overall system efficiency \cite{liu2021vehicular, murshed2021machine}. This paradigm is particularly advantageous for applications requiring real-time data processing, such as autonomous vehicles, healthcare systems, and industrial automation, where rapid decision-making and immediate data analysis are essential. However, despite its benefits, edge computing alone often faces limitations in terms of resources and scalability \cite{du2023computation}.

To overcome these challenges, the Edge-Cloud Continuum integrates both edge and cloud resources, allowing more intensive computational tasks to be handled in the cloud while maintaining low-latency operations at the edge. Despite its potential, efficiently placing applications within this continuum presents significant challenges. One of the primary difficulties is ensuring that applications meet Service Level Agreements (SLAs), which set specific performance and availability metrics. Failing to meet SLAs can result in severe penalties and a degraded Quality of Service (QoS) for end-users \cite{wang2019towards, faticanti2020, konzen2023}.

Resource sharing among providers within an edge-cloud continuum environment allows them to offer sufficient resources to provision applications, thus reducing SLA violations and improving both QoS and Quality of Experience (QoE) for end-users \cite{xia2020qoe, kochovski2020smart}. However, the dynamic and heterogeneous nature of these environments makes the resource allocation process more complex. Variations in demand, coupled with the diverse capabilities of edge and cloud resources, require advanced, real-time strategies for application placement to optimize resource use and meet SLAs \cite{apat2023comprehensive, alwabel2024deadline}. Application placement involves allocating modules or components of applications across computing resources in a manner that minimizes disruption and optimizes resource use \cite{qayyum2018fognetsim++}, a requirement for sustaining edge-cloud applications that demand low-latency responsiveness and real-time performance \cite{nandhakumar2024}.

Current solutions attempt to address the application placement problem by developing allocation plans that balance user demands—such as minimizing delay and latency—with provider-side infrastructure constraints \cite{apat2023comprehensive}. Several strategies rely on heuristic approaches, reinforcement learning, and game-theoretic models to optimize placement decisions. For instance, latency-aware and energy-efficient scheduling mechanisms have been explored to improve computational efficiency \cite{alwabel2024deadline}. Additionally, federated edge-cloud infrastructures have been studied to enhance scalability and resource sharing \cite{faticanti2020}. However, these approaches often focus on static or semi-dynamic resource allocation, which limits their applicability in highly dynamic environments. Moreover, most methods prioritize either latency minimization or SLA compliance but fail to achieve an effective trade-off between both objectives.

Despite these advancements, existing solutions still struggle to efficiently manage dynamic workloads in heterogeneous edge-cloud environments. Unpredictable fluctuations in demand, variations in network conditions, and constraints on computational resources pose significant challenges to application placement strategies. Current methods frequently overlook the importance of adaptive mechanisms that can dynamically reallocate resources in response to changing conditions. Furthermore, many approaches neglect the impact of application drops, which can critically degrade the end-user experience. Therefore, there is a need for an application placement strategy that effectively balances SLA compliance, latency minimization, and application reliability while dynamically adjusting to workload fluctuations.

To address these limitations, we propose Tetris, an edge-cloud continuum application placement strategy inspired by the iconic puzzle game. Similar to the game, where the objective is to fit pieces together to complete lines and prevent buildup, the Tetris approach strategically allocates computational services across edge-cloud resources to optimize SLA compliance and resource utilization. Our approach employs a heuristic algorithm that prioritizes services based on urgency and resource availability, effectively minimizing SLA violations and preventing system overload.

This paper presents three main contributions:

\begin{enumerate}
    \item A novel SLA-aware heuristic algorithm for application placement in the edge-cloud continuum, which dynamically prioritizes tasks based on urgency, workload fluctuations, and available resources.
    \item An adaptive resource management strategy that minimizes SLA violations and optimizes resource utilization by dynamically redistributing services across edge and cloud nodes.
    \item A comprehensive performance evaluation, demonstrating that Tetris significantly reduces application drops and SLA violations compared to state-of-the-art methods.
\end{enumerate}

The remainder of this paper is structured as follows: Section \ref{sec:rw} presents the related works, and Section \ref{sec:sm} highlights the System Model adopted in this work. Section \ref{sec:tetris} describes the proposed strategy. The details of the simulation setup are provided in Section \ref{sec:experiments}. Section \ref{sec:evaluation} analyzes the results, and Section \ref{sec:conclusion} concludes the study.

\section{Related Work}
\label{sec:rw}

Several studies have explored different strategies to optimize these processes. The works in \cite{rw_extra1_multi} and \cite{rw_extra2_energy} employed deadline-aware, energy-aware, and latency-aware mechanisms to enhance placement and scheduling in fog-computing environments. In \cite{rw_extra3_ecc}, the focus shifted to an edge-cloud continuum scenario, where placement and scheduling were analyzed. Similarly, \cite{rw_extra4_cmfog}, \cite{9428571}, and \cite{LIX202495OLIVEIRA} introduced solutions tailored for multi-level fog infrastructures. Meanwhile, \cite{souza2022b} and \cite{souzathea2023} proposed heuristics to allocate computational services in edge computing under SLA constraints.

One relevant approach is the Communication-Based Edgewards (CB-E) scheduler \cite{9428571}, which dynamically allocates service components across fog layers to meet latency requirements. Evaluation results demonstrate its effectiveness in improving responsiveness in fog computing. Another notable strategy is the Least Impact - X (LI-X) algorithm \cite{LIX202495OLIVEIRA}, which optimizes latency-sensitive applications by strategically distributing modules across fog hierarchies, reducing network traffic while maintaining low response times.

In the domain of SLA-aware resource allocation, Argos \cite{souza2022b} was proposed to ensure compliance with latency constraints while preserving privacy-sensitive applications in edge environments. By intelligently migrating tasks across edge servers, Argos demonstrated a strong ability to balance performance and security.

A more recent and promising approach is Thea \cite{souzathea2023}, which directly tackles the challenge of provisioning services on edge computing servers while ensuring sufficient capacity, reliability, and SLA compliance. Thea adopts a resource-aware placement strategy that dynamically adjusts allocation to minimize SLA violations. Experimental results highlight Thea’s ability to significantly outperform prior approaches in reducing SLA breaches, reinforcing its position as one of the most effective state-of-the-art solutions for dynamic edge environments.

An overview of the works mentioned in this section is presented in Table \ref{tab:compare_rw}.

\begin{table}[ht!]
\centering
\setlength{\tabcolsep}{4pt}
\renewcommand{\arraystretch}{1.4}
\caption{Comparison between Tetris and related works. (1) Fog-Edge Only, (2) Edge-Cloud Continuum, (3) Modular App, (4) Latency, (5) Drop, (6) Power, (7) Deadline.}
\label{tab:compare_rw}

\def\cellZ{\cellcolor{black!10}\xmark}
\def\cellO{\cmark}

\begin{tabular}{>{\raggedright\arraybackslash}p{2.7cm} >{\centering\arraybackslash}p{0.55cm} >{\centering\arraybackslash}p{0.325cm} >{\centering\arraybackslash}p{0.325cm} >{\centering\arraybackslash}p{0.325cm} >{\centering\arraybackslash}p{0.325cm} >{\centering\arraybackslash}p{0.325cm} >{\centering\arraybackslash}p{0.325cm} >{\centering\arraybackslash}p{0.325cm}}
    \toprule
    \multirow{2}{*}{\textbf{Work}} & \multirow{2}{*}{\textbf{Year}} & \multicolumn{7}{c}{\textbf{Features}} \\ 
    \cmidrule(lr){3-9}
    & & \textbf{1} & \textbf{2} & \textbf{3} & \textbf{4} & \textbf{5} & \textbf{6} & \textbf{7} \\ 
    \midrule
    LAMOMRank \cite{rw_extra1_multi}    & 2024 & \cellO{} & \cellZ{} & \cellO{} & \cellO{} & \cellZ{} & \cellO{} & \cellZ{} \\ 
    PEAPM/POAPM \cite{rw_extra2_energy} & 2024 & \cellZ{} & \cellO{} & \cellO{} & \cellO{} & \cellZ{} & \cellO{} & \cellO{} \\ 
    LIP \cite{rw_extra3_ecc}            & 2024 & \cellZ{} & \cellO{} & \cellO{} & \cellO{} & \cellZ{} & \cellZ{} & \cellZ{} \\ 
    CMFog$_{V}$ \cite{rw_extra4_cmfog}  & 2024 & \cellZ{} & \cellO{} & \cellO{} & \cellO{} & \cellZ{} & \cellZ{} & \cellZ{} \\ 
    CB-E \cite{9428571}                 & 2022 & \cellZ{} & \cellO{} & \cellO{} & \cellO{} & \cellZ{} & \cellZ{} & \cellZ{} \\ 
    LI-X \cite{LIX202495OLIVEIRA}       & 2024 & \cellZ{} & \cellO{} & \cellO{} & \cellO{} & \cellZ{} & \cellZ{} & \cellZ{} \\ 
    Argos \cite{souza2022b}             & 2022 & \cellO{} & \cellZ{} & \cellO{} & \cellO{} & \cellZ{} & \cellZ{} & \cellO{} \\ 
    Thea \cite{souzathea2023}           & 2023 & \cellO{} & \cellZ{} & \cellO{} & \cellO{} & \cellZ{} & \cellO{} & \cellO{} \\ 
    \textbf{Tetris}                     & 2025    & \cellO{} & \cellO{} & \cellO{} & \cellO{} & \cellO{} & \cellO{} & \cellO{} \\ 
    \bottomrule
\end{tabular}
\end{table}

Table \ref{tab:compare_rw} provides a comparative analysis of scheduling and placement strategies in fog and edge computing environments. Thea \cite{souzathea2023}, which demonstrated superior performance over Argos \cite{souza2022b}, is recognized as an effective SLA-aware placement approach, optimizing resource allocation while ensuring reliability. However, its scope remains confined to fog and edge infrastructures, without considering the edge-cloud continuum. Additionally, Thea does not account for drop as a service degradation metric, limiting its ability to model performance fluctuations due to resource contention and network instability. In contrast, Tetris integrates latency, drop, power consumption, and deadline as core optimization criteria, ensuring a more comprehensive and adaptive placement strategy. Therefore, Tetris enhances placement decisions across multi-tier infrastructures, addressing SLA compliance.

\section{System Model}
\label{sec:sm}

\begin{figure}[ht!]
    \centering
    \includegraphics[width=7.25cm]{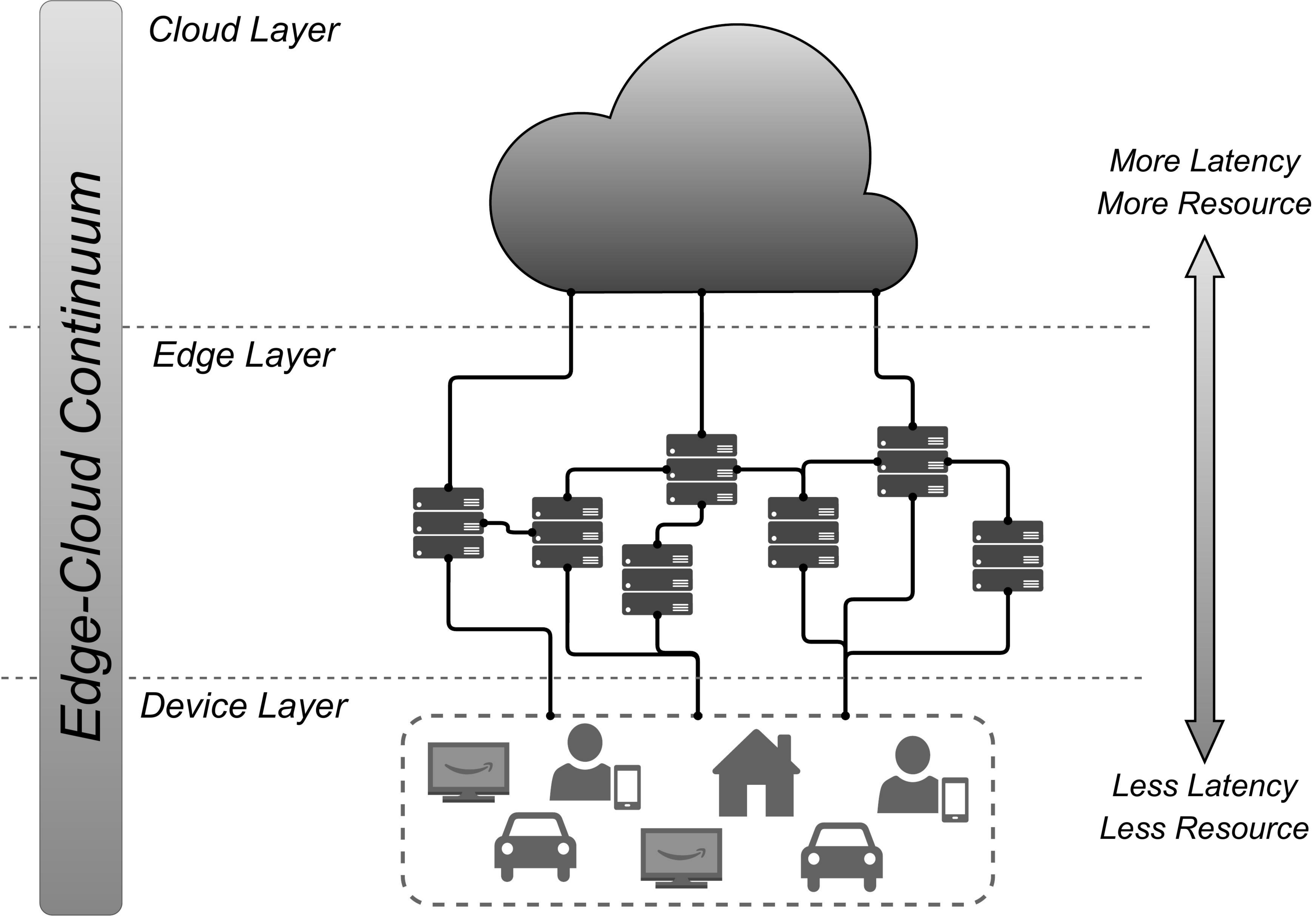}
    \caption{Edge–Cloud Continuum Environment}
    \label{fig:overview}
\end{figure}

Consider an edge-cloud continuum represented as a graph $G = (V, L)$, where the node set \(V\) includes end devices, edge servers, and cloud servers, and \(L\) denotes the set of network links connecting these nodes. Each node \(v \in V\) is defined by a resource vector $\mathbf{R}_v = \bigl(C_v, M_v, S_v\bigr)$, representing its computational capacity, memory, and storage, in addition to network parameters such as latency and effective bandwidth. The architecture, depicted in Figure~\ref{fig:overview}, is organized hierarchically: end devices at the base, edge servers in an intermediate layer, and cloud servers at the top. This structure captures the intrinsic trade-off between resource availability and communication latency.

In this environment, users \(\mathcal{U}\) generate a stream of tasks \(\mathcal{T}\), where each task \(t \in \mathcal{T}\) is specified by a resource requirement vector $\mathbf{r}_t = \bigl(r_t^{\mathrm{CPU}}, r_t^{\mathrm{RAM}}, r_t^{\mathrm{Storage}}\bigr)$, an estimated processing time \(T_t\), an input data size \(\delta_t\), an arrival time \(a_t\), a deadline \(d_t\), and a penalty coefficient \(\lambda_t\) for SLA violations. The operational model is defined by two interdependent phases: a scheduling phase that determines the temporal execution order of tasks and a placement phase that allocates the scheduled tasks to nodes in the continuum.

In the scheduling phase, a composite metric \(\phi(t)\) is computed for each task to aggregate its SLA-critical parameters. This metric is formulated as a weighted sum of the deadline \(d_t\), the processing time \(T_t\), and the communication delay \(\delta_t/B_{us}\), where \(B_{us}\) denotes the effective bandwidth between the user and a prospective processing node. The start time \(S_t\) is determined based on the task’s arrival and the accumulated delays, so that the finish time is given by $F_t = S_t + T_t + \frac{\delta_t}{B_{us}}$. Any excess over the allowable deadline \(a_t + d_t\) is penalized by a term \(\lambda_t\,\max\{0,\,F_t - (a_t+d_t)\}\). Tasks are thus ordered in a non-decreasing order of \(\phi(t)\), ensuring that those with more stringent temporal requirements are prioritized.

Following scheduling, the placement phase allocates tasks to nodes while minimizing resource fragmentation and satisfying capacity constraints. Fragmentation occurs when resources are unevenly distributed across nodes, leading to inefficient utilization and limiting the ability to accommodate future tasks. To address this, each node \( v \in V \) is evaluated using a composite capacity metric $\gamma(v) = \Bigl( R_{v,\mathrm{free}}^{\mathrm{CPU}} R_{v,\mathrm{free}}^{\mathrm{RAM}} R_{v,\mathrm{free}}^{\mathrm{Storage}} \Bigr)^{\frac{1}{3}}$, where \( R_{v,\mathrm{free}}^{\cdot} \) represents the residual capacity of each resource after accounting for previously allocated tasks. This metric prioritizes nodes with a more balanced availability of CPU, memory, and storage, reducing fragmentation.

Tasks, already prioritized by the scheduling phase, are iteratively assigned to the node that minimizes \( \gamma(v) \) while meeting the condition \( \mathbf{r}_t \leq \mathbf{R}_v \). This approach ensures that tasks are allocated to nodes with the lowest residual imbalance, maintaining available capacity in a way that facilitates future allocations. Additionally, the placement must satisfy the constraint that the cumulative resource demands on any node do not exceed its capacity, and that total data transfer over each network link remains within its bandwidth \( B_{vw} \).

The integrated optimization problem is simplified as follows. Our goal is to minimize the overall cost incurred by SLA violations, latency, task drops, and energy consumption. In a standard optimization formulation, the problem is stated as:

\[
\min_{S_t,\, F_t,\, D_t,\, E_t} J = \sum_{t \in \mathcal{T}} \bigl[ \lambda_t\,\Delta_t + \omega (F_t - a_t) + \rho D_t + \eta E_t \bigr].
\]

\begin{equation}
\substack{\mathsf{Resources} \\ \mathsf{Capacity}}
\left\{
\begin{aligned}
&\sum_{t \in \mathcal{T}_v} r_t^{\mathrm{CPU}} \le C_v, \quad &&\forall\, v \in V,\\
&\sum_{t \in \mathcal{T}_v} r_t^{\mathrm{RAM}} \le M_v, \quad &&\forall\, v \in V,\\
&\sum_{t \in \mathcal{T}_v} r_t^{\mathrm{Storage}} \le S_v, \quad &&\forall\, v \in V
\end{aligned}
\right.
\tag{1}
\end{equation}

\begin{equation}
\substack{\mathsf{Network}}
\left\{
\begin{aligned}
\sum_{t \in \mathcal{T}_{vw}} \frac{\delta_t}{B_{vw}} \leq L_{vw}^{\max}, \quad \forall\, (v,w) \in L
\end{aligned}
\right.
\tag{2}
\end{equation}

\begin{equation}
\substack{\mathsf{Scheduling}}
\left\{
\begin{aligned}
F_t = S_t + T_t + \frac{\delta_t}{B_{us}}, \quad \forall\, t \in \mathcal{T},
\end{aligned}
\right.
\tag{3}
\end{equation}

\begin{equation}
\substack{\mathsf{Deadline}} 
\left\{
\begin{aligned}
\Delta_t = \max\{0,\,F_t - (a_t+d_t)\}, \quad \forall\, t \in \mathcal{T}\end{aligned}
\right.
\tag{4}
\end{equation}

\vspace{0.5cm}

The constraints are as follows:

\begin{enumerate}
    \item \textbf{Resource Capacity Constraints:} For each node \(v \in V\), the cumulative resource demands of all tasks allocated to \(v\) must not exceed its available capacities:
    
    \item \textbf{Network Constraints:} For each network link \((v,w) \in L\), the total communication delay incurred by tasks whose data traverse the link must not exceed the maximum tolerable delay:
    
    \item \textbf{Scheduling Constraint:} The finish time \(F_t\) for each task \(t\) is determined by its start time \(S_t\), processing time \(T_t\), and communication delay \(\delta_t/B_{us}\):
    
    \item \textbf{Deadline Violation:} The variable \(\Delta_t\) quantifies the delay beyond the deadline \(a_t+d_t\) and is defined as

\end{enumerate}

In this formulation:
\begin{itemize}
    \item \(\lambda_t\,\Delta_t\) penalizes any delay beyond the deadline \(a_t+d_t\), where $\Delta_t = \max\{0,\,F_t - (a_t+d_t)\}$.
    \item \(\omega\,(F_t - a_t)\) represents the total latency for task \(t\).
    \item \(\rho\,D_t\) is the cost incurred when task \(t\) is dropped due to insufficient resources.
    \item \(\eta\,E_t\) quantifies the energy consumption cost for processing task \(t\).
    \item The coefficients \(\lambda_t\), \(\omega\), \(\rho\), and \(\eta\) balance the importance of deadline adherence, latency minimization, reliability, and energy efficiency.
\end{itemize}

This formulation represents the dual processes of scheduling and placement in a dynamic edge-cloud environment. The scheduling phase establishes a temporal order that minimizes SLA penalties by prioritizing tasks with critical deadlines and significant communication overhead, while the placement phase assigns these tasks to nodes in a way that optimizes resource utilization under capacity and network constraints. Table~\ref{tab:symbols} presents the key symbols and notations used throughout this paper.

\begin{table}[ht!]
\centering
\scriptsize
\setlength{\tabcolsep}{4pt}
\renewcommand{\arraystretch}{1.18}
\caption{List of Symbols}
\label{tab:symbols}
\begin{tabular}{p{1.7cm} p{6.1cm}}
\hline
\textbf{Symbol} & \textbf{Description} \\
\hline
\(G = (V,L)\) & Graph representing the edge-cloud continuum \\
\(V\) & Set of nodes (end devices, edge servers, cloud servers) \\
\(L\) & Set of network links \\
\(\mathbf{R}_v\) & Resource capacity vector of node \(v\): CPU, memory, storage \\
\(R\) & Residual capacities of node \(v\) after allocation \\
\(\mathcal{U}\) & Set of users \\
\(\mathcal{T}\) & Set of tasks generated by users \\
\(\mathbf{r}_t\) & Resource requirement vector of task \(t\) \\
\(T_t\) & Estimated processing time of task \(t\) \\
\(\delta_t\) & Input data size of task \(t\) \\
\(a_t\) & Arrival time of task \(t\) \\
\(d_t\) & Deadline of task \(t\) \\
\(\lambda_t\) & Penalty coefficient for deadline violations of task \(t\) \\
\(S_t\) & Start time of task \(t\) \\
\(F_t\) & Finish time of task \(t\) \(\left(F_t = S_t + T_t + \frac{\delta_t}{B_{us}}\right)\) \\
\(B_{us}\) & Effective bandwidth between the user of task \(t\) and a processing node \\
\(\phi(t)\) & Composite SLA metric for task \(t\) \(\left(\phi(t) = \alpha\,d_t + \beta\,T_t + \gamma\,\frac{\delta_t}{B_{us}}\right)\) \\
\(\gamma(v)\) & Composite capacity metric for node \(v\) \(\left(\gamma(v) = \Bigl( R_{v,\mathrm{free}}^{\mathrm{CPU}}\,R_{v,\mathrm{free}}^{\mathrm{RAM}}\,R_{v,\mathrm{free}}^{\mathrm{Storage}} \Bigr)^{\frac{1}{3}}\right)\) \\
\(\omega,\, \rho,\, \eta\) & Weighting coefficients for latency, drop cost, and energy consumption, respectively \\
\(D_t\) & Cost associated with dropping task \(t\) \\
\(E_t\) & Energy consumption cost for executing task \(t\) \\
\(\Delta_t\) & Deadline violation of task \(t\) \(\left(\Delta_t = \max\{0,\,F_t - (a_t+d_t)\}\right)\) \\
\(B_{vw}\) & Bandwidth of network link \((v,w)\) \\
\(L_{vw}^{\max}\) & Maximum tolerable aggregated communication delay over link \((v,w)\) \\
\hline
\end{tabular}
\end{table}


\section{Tetris Design}
\label{sec:tetris}

Tetris is a placement algorithm designed for latency-sensitive applications within the Edge-Cloud Continuum. These applications have varying deadlines, and the service provider is required to meet these deadlines as defined in the SLA (Service Level Agreement). Failure to comply with the SLA results in penalties, including costs related to the execution of tasks and increased energy consumption. To meet these requirements, Tetris prioritizes tasks with shorter deadlines, considering the computational resources available on edge and cloud servers to avoid task drops and reduce latency.

\begin{figure}[h]
	\centering
	\includegraphics[width=6.5cm]{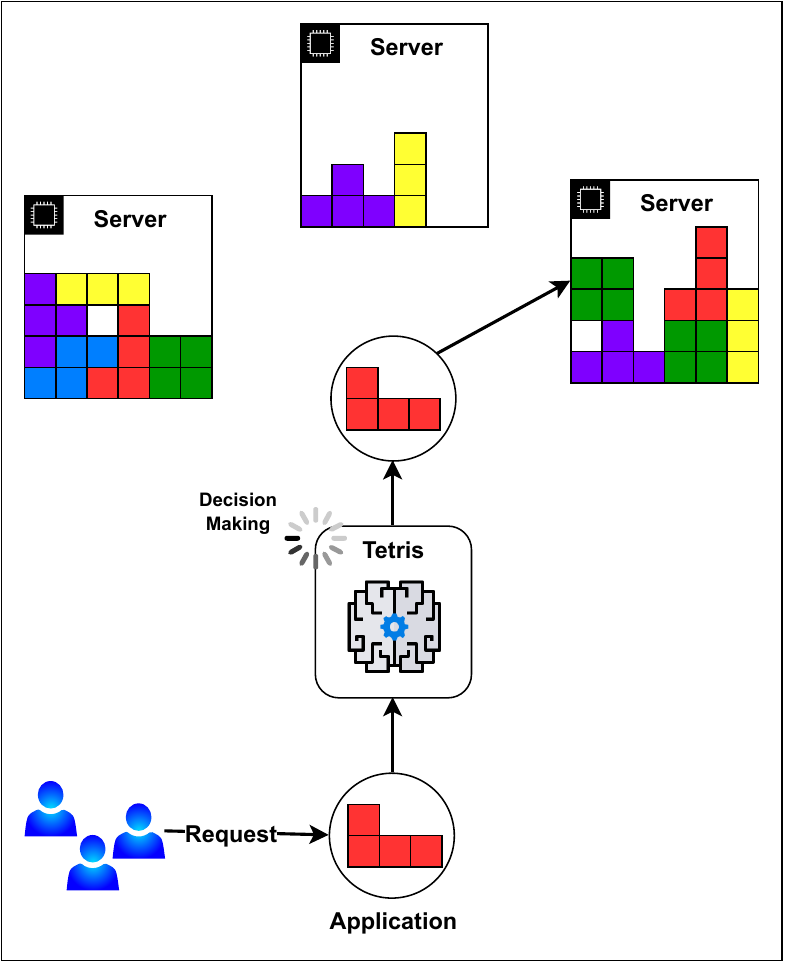}
	\caption{Overview of Tetris algorithm concept}
	\label{fig:tetris}
\end{figure}

Figure \ref{fig:tetris} illustrates the placement process. Users request services, and after receiving the request, Tetris makes placement decisions based on the available server resources and the SLA requirements. The algorithm selects the most appropriate server to allocate each task. This process combines two strategies: one focused on minimizing SLA violations, as inspired by the approach in \cite{faticanti2020}, and the other on optimizing resource use through the Most Capacity First algorithm, as proposed in \cite{lai2020cost}.

The Tetris algorithm begins by initializing two lists: a list of tasks to be allocated (\(\mathcal{T}\)) and a list of servers (\(V\)). An empty list, \textit{SLAList}, is created to store the metrics related to potential SLA violations for each task. After calculating the SLA violation metrics for all tasks in \(\mathcal{T}\), i.e., \(\phi(t)\), the algorithm sorts the tasks in ascending order based on these metrics. The processing capacity of each server is then evaluated, considering available CPU, RAM, and disk space. This information is stored in the \textit{NodeList}, which is also sorted in ascending order. 

The metric $\phi(t) = \alpha\,d_t + \beta\,T_t + \gamma\,\frac{\delta_t}{B_{us}}$, integrates the task's deadline, processing time, and communication delay into a single value. This formulation prioritizes tasks with critical deadlines and high communication overhead, ensuring that SLA-sensitive tasks are addressed first. The $\gamma(v)$ captures the balanced availability of resources at node \(v\). By employing the geometric mean, the algorithm effectively de-emphasizes nodes with imbalanced resource distributions, thereby mitigating resource fragmentation and facilitating the accommodation of future tasks.

For each node \(v \in V\) in the ordered list of servers, the algorithm checks if \(v\) has the necessary capacity to host the task \(t \in \mathcal{T}\) (i.e., if \(\mathbf{r}_t \leq \mathbf{R}_v\)). If the node has sufficient resources and the task is not yet provisioned on \(v\), the algorithm assigns \(t\) to \(v\). This process ensures that tasks are allocated efficiently across available servers, maximizing resource utilization.

\begin{algorithm}[ht!]
    \small
	\caption{Tetris Algorithm} 
	\begin{algorithmic}[1]
		\State \(\mathcal{T} \leftarrow\) List of Tasks   
		\State \(V \leftarrow\) List of Servers
		\State \textit{SLAList} \( \leftarrow \{\} \) 
		\For {each task \(t \in \mathcal{T}\)}
			\State \textit{SLAList}[\(t\)] \( \leftarrow \phi(t) \)
		\EndFor
		\State \(\mathcal{T}' \leftarrow\) sortAscending(\textit{SLAList})    
		\For {each task \(t \in \mathcal{T}'\)}
			\State \textit{NodeList} \( \leftarrow \{\} \) 
			\For {each node \(v \in V\)}
				\State \textit{NodeList}[\(v\)] \( \leftarrow \Bigl( R_{v,\mathrm{free}}^{\mathrm{CPU}} \times R_{v,\mathrm{free}}^{\mathrm{RAM}} \times R_{v,\mathrm{free}}^{\mathrm{Storage}} \Bigr)^{\frac{1}{3}} \)
			\EndFor
			\State \(V' \leftarrow\) sortAscending(\textit{NodeList})
			\For {each node \(v \in V'\)}
				\If{\( \mathbf{r}_t \leq \mathbf{R}_v \)}
					\If{\(t\) is not provisioned on \(v\)} 
						\State Provision \(t\) on \(v\)
					\EndIf
					\State \textbf{break}
				\EndIf
			\EndFor
		\EndFor
	\end{algorithmic}
	\label{algo}
\end{algorithm}

This tight integration between the scheduling phase, which prioritizes tasks based on \(\phi(t)\), and the placement phase, which selects nodes according to \(\gamma(v)\), ensures that tasks with stringent SLA requirements are allocated to nodes that can best meet their resource demands. As a result, overall system performance is optimized by minimizing both SLA violations and resource fragmentation. Therefore, the Tetris algorithm addresses the challenge of allocating tasks to edge-cloud servers by focusing on both SLA compliance and resource efficiency. By balancing these two factors, Tetris offers a near-optimal task placement, minimizing SLA violations and avoiding task drops. The following section provides a performance evaluation of the algorithm.



\begin{figure}[ht!]
	\centering
	\includegraphics[width=6.5cm]{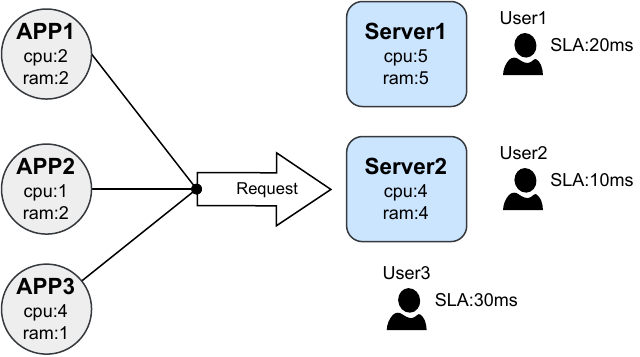}
	\caption{Allocation Request}
	\label{fig:toyrequest}
\end{figure}

To illustrate the allocation process and analyze the decision-making behavior under controlled conditions, we consider a toy example. Figure~\ref{fig:toyrequest} shows the allocation request. Three applications denoted as APP1, APP2, and APP3, are considered with resource requirements as follows: APP1 requires 2 CPU units and 2 RAM units; APP2 requires 1 CPU unit and 2 RAM units; APP3 requires 4 CPU units and 1 RAM unit. Two servers are available: Server1 with 5 CPU units and 5 RAM units, and Server2 with 4 CPU units and 4 RAM units. The associated SLA's (deadlines for the applications) are 20ms, 10ms, and 30ms, respectively.

\begin{figure}[ht!]
	\centering
	\includegraphics[width=5.75cm]{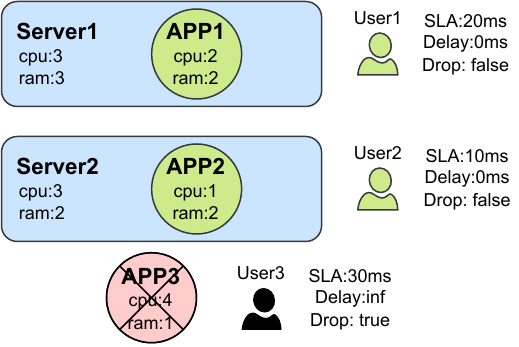}
	\caption{Toy example of Thea algorithm}
	\label{fig:toythea}
\end{figure}

Figure~\ref{fig:toythea} shows the placement outcome using the Thea algorithm. In this allocation, the algorithm chooses the shortest path between the user and the server that will host the user's application. This way, Thea can minimize the application delivery delay. This effect is evident as APP1 and APP2 are processed with zero delay, whereas APP3 receives an infinite delay, resulting in a task drop. The failure to allocate APP3 is due to resource fragmentation, which prevents the formation of a contiguous block of resources necessary to host the application. Therefore, while Thea is able to minimize delay, in environments with limited resources, this behavior creates resource fragmentation, leading to SLA violations.

\begin{figure}[ht!]
	\centering
	\includegraphics[width=5.75cm]{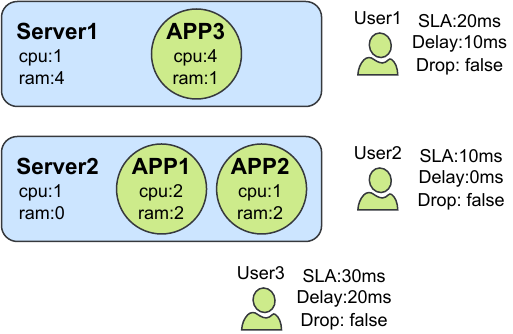}
	\caption{Toy example of Tetris algorithm}
	\label{fig:toytetris}
\end{figure}

Figure~\ref{fig:toytetris} presents the placement outcome using the Tetris algorithm. In this result, the algorithm first creates an application queue based on deadlines (Earliest Deadline First) and then chooses the server with the lowest capacity first. This way, Tetris is able to avoid resource fragmentation and becomes more resilient in terms of task drops. In the toy example, this effect is clear: Tetris can arrange and deploy all applications using the available resources more smartly without violating the users' SLAs. Although the average delay is greater than with Thea, there are no task drops, and all users' requests are fulfilled.

Note that the integration between the scheduling phase (which leverages \(\phi(t)\)) and the placement phase (which utilizes \(\gamma(v)\)) is clearly demonstrated in this toy example. For instance, consider APP3: due to its stringent SLA requirements—reflected in a high \(\phi(t)\) value—it is prioritized for early allocation. However, when evaluating the available nodes using \(\gamma(v)\), the algorithm identifies that none of the initially chosen nodes can accommodate APP3 without causing fragmentation, leading to a task drop in the case of Thea. In contrast, Tetris dynamically recalculates the resource availability, selecting an alternative node with a more balanced resource profile (i.e., a lower \(\gamma(v)\)) that is capable of hosting APP3. This dynamic adjustment ensures that even tasks with critical deadlines are matched with nodes that best meet their resource demands.

Furthermore, in scenarios with high variability, periodic updates to both \(\phi(t)\) and \(\gamma(v)\) allow Tetris to adapt in real-time to changes in workload or network conditions. This tight integration between scheduling and placement minimizes SLA violations and enhances overall system resilience by reducing task drops and preventing resource fragmentation. In this manner, Tetris aligns task urgency with the selection of nodes exhibiting resource balance, resulting in improvements in both SLA adherence and overall resource efficiency.


\section{Experiment Setup}
\label{sec:experiments}

\begin{figure*}[ht!]
	\centering
	\includegraphics[width=13cm]{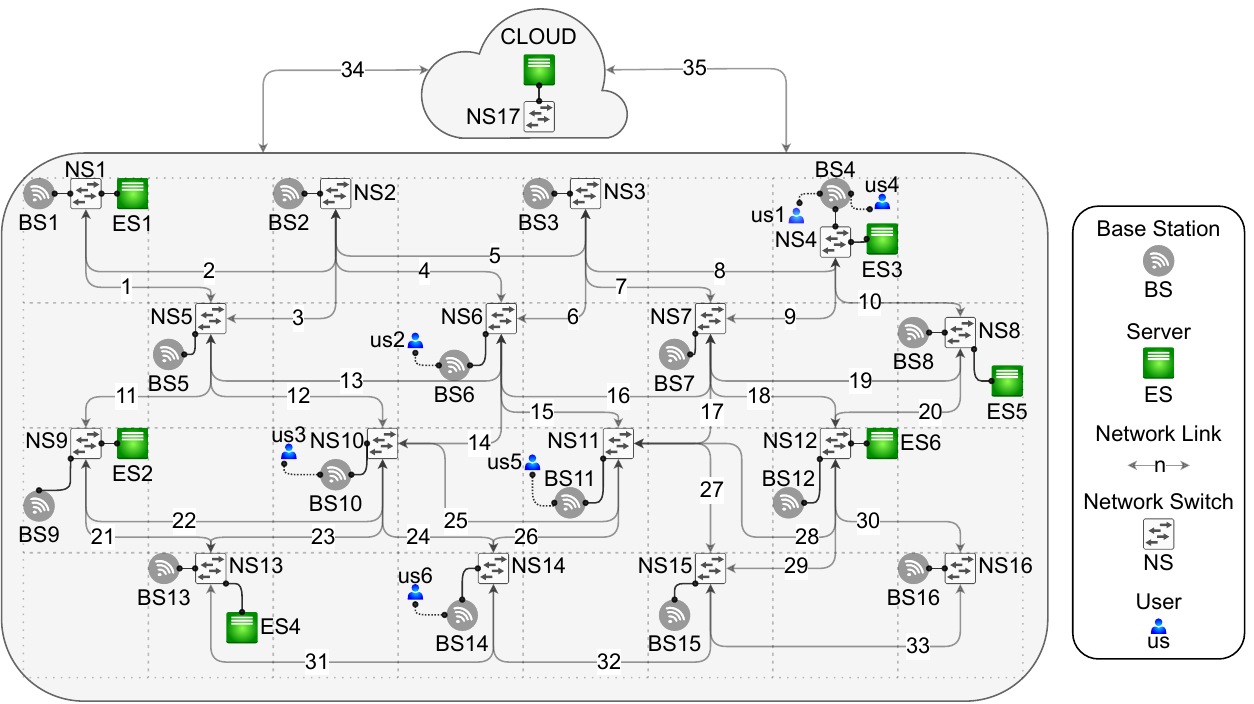}
	\caption{Experimental Scenario: The setup used in the simulation comprised 7 servers (6 edge servers labeled ES1 to ES6 and 1 cloud server labeled CLOUD) connected via 35 network links and 17 network switches (labeled NS1 to NS17). These switches connect the edge servers to 16 base stations (BS1 to BS16), serving as access points for 6 users (us1 to us6). Users are connected to specific base stations: us1 and us4 to BS4, us2 to BS6, us3 to BS10, us5 to BS11, and us6 to BS14.}
	\label{fig:scenario}
\end{figure*}

The EdgeSimPy simulator (v1.1.0) \cite{souza2023} was used to evaluate resource allocation strategies in edge computing. To facilitate reproducibility and sharing, an edge-cloud continuum scenario was proposed based on a dataset available in the simulator (sample\_dataset2.json). This scenario is illustrated in Figure \ref{fig:scenario}.

The infrastructure includes 7 servers connected through 35 network links to 17 network switches. These servers are categorized into 6 edge servers and 1 cloud server, with specifications in Table \ref{tab:server_parameters}. The network switches link edge servers to base stations, which serve as access points for 6 users. A 1:1 relationship was defined between users and applications, with users distributed across different base stations. Users remained stationary during the simulation.

\begin{table} [ht!]
\small
\centering
  \caption{Servers Specification.}
  \label{tab:server_parameters}
  \begin{tabular}{cccc}
    \toprule
    Server & CPU & Memory & Disk\\
    \midrule
    CLOUD & 9 $\cdot 10^{3}$ &  9 $\cdot 10^{6}$  & 9 $\cdot 10^{7}$ \\ 
    ES1 \& ES2 & 8 &  1.6 $\cdot 10^{4}$  & 1.3 $\cdot 10^{5}$ \\ 
    ES3 \& ES4 & 8 & 8 $\cdot 10^{3}$ & 1.3 $\cdot 10^{5}$ \\ 
    ES5 \& ES6  & 12 &  1.6 $\cdot 10^{4}$  & 1.3 $\cdot 10^{5}$ \\
  \bottomrule
\end{tabular}
\end{table}

Using this edge-cloud continuum scenario, the Tetris algorithm was compared to Thea \cite{souzathea2023}, a state-of-the-art algorithm focused on distributing computational services across server resources. However, Thea has not been evaluated in edge-cloud scenarios or high workload situations and does not consider application drop behavior. This study addresses these aspects. Tetris distinguishes itself by not merely seeking the fastest path or nearest server; instead, it employs a smart decision-making process that mitigates overloading and prevents application drops, particularly in high workload scenarios.

\subsection{Experiment Planning}

To evaluate the performance of strategies in the edge-cloud continuum context, a complete factorial design \cite{jain1991art} was used. This approach measures Latency SLA Violations, Drop SLA Violations, Average Latency for delivering services, and Power Consumption by considering all combinations of factors and their levels. Each of the three factors (Algorithm, Workload, and Cloud) has two levels, resulting in \textbf{$2^3$} combinations. Table \ref{tab:exp_factors} shows the factors and their respective levels.

\begin{table}[ht!]
\small
\centering
  \caption{Factors and Levels.}
  \label{tab:exp_factors}
  \begin{tabular}{cc|c}
    \toprule
    Factors & \multicolumn{2}{c}{Levels} \\
    \midrule
    Algorithm & Tetris & Thea  \\ 
    Workload & Low & High \\ 
    Cloud & On & Off \\ 
  \bottomrule
\end{tabular}
\end{table}

For Workload, the Low level corresponds to applications with CPU demands ranging from 2 to 8 cores and memory demands from 2048 to 8192 GB. The High level has CPU values between 6 and 12 cores and memory ranging from 2048 to 16384 GB. For Cloud, "On" means the cloud server is active (Edge-Cloud), and "Off" means the cloud server is not available (Edge-Only).

Next, we define the response variables to clarify their roles in the evaluation. Latency SLA Violations represent the number of times the algorithm failed to deliver the requested service to the user within the agreed time. Drop SLA Violations indicate the number of application drops during the simulation. An application is considered dropped when the simulation ends and users are still waiting for the application. At the same time, a latency violation occurs when the user receives the requested application after the agreed deadline but before the simulation finishes. Although Latency SLA Violations signify a degradation of QoS, Drop SLA Violations are more severe because the users do not receive the application. In addition to these violations, Average Latency is the mean time taken to complete the entire process, from the user's request arrival to the time the user accesses the service. Power Consumption refers to the energy consumed by the servers during the simulation.

Each combination was replicated ten times with different seeds. A 95\% confidence interval was defined for the analysis of results, allowing data dispersion intervals to be added to bar graphs. The values for each response variable were calculated as their mean and standard deviation, along with the influence of each factor on their behavior.

\section{Result Analysis}
\label{sec:evaluation}

\begin{figure}[ht!]
    \centering
    \scriptsize
    \begin{tikzpicture}
        \begin{axis}[
            ybar,
            symbolic x coords={Algorithm, Workload, Cloud, {Algorithm\\Workload}, {Algorithm\\Cloud}, {Workload\\Cloud}, {Algorithm\\Workload\\Cloud}},
            xtick=data,
            ymin=0,
            ymax=50,
            ylabel={Percentage Influence (\%)},
            bar width=5pt,
            enlarge x limits=0.05,
            xticklabel style={align=center},
            width=9cm,
            height=5cm,
            legend style={at={(0.5,1.2)}, anchor=north, legend columns=3},
            grid=major,
            xtick distance=1
        ]
        \addplot[ybar, fill=black!80, draw=black] coordinates {
            (Algorithm, 20.83)
            (Workload, 12.99)
            (Cloud, 13.45)
            ({Algorithm\\Workload}, 10.79)
            ({Algorithm\\Cloud}, 12.87)
            ({Workload\\Cloud}, 14.86)
            ({Algorithm\\Workload\\Cloud}, 14.20)
        };
        \addplot[ybar, fill=black!50, draw=black] coordinates {
            (Algorithm, 24.63)
            (Workload, 3.36)
            (Cloud, 43.40)
            ({Algorithm\\Workload}, 0.16)
            ({Algorithm\\Cloud}, 25.83)
            ({Workload\\Cloud}, 2.41)
            ({Algorithm\\Workload\\Cloud}, 0.22)
        };
        \addplot[ybar, fill=black!10, draw=black] coordinates {
            (Algorithm, 8.96)
            (Workload, 10.44)
            (Cloud, 32.52)
            ({Algorithm\\Workload}, 40.10)
            ({Algorithm\\Cloud}, 4.44)
            ({Workload\\Cloud}, 3.52)
            ({Algorithm\\Workload\\Cloud}, 0.01)
        };
        \legend{Latency SLA Violations, Average Latency, Power Consumption}
        \end{axis}
    \end{tikzpicture}
    \caption{Influence of each factor on Latency SLA Violations, Average Latency, and Power Consumption}
    \label{fig:influence_combined}
\end{figure}
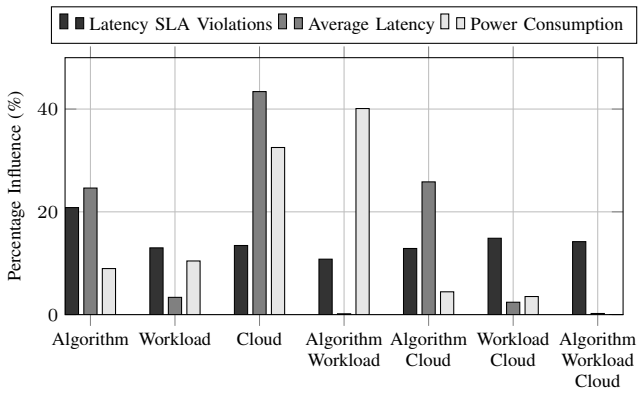

The influence plot, illustrated in Figure \ref{fig:influence_combined}, show the impact of various factors on the response variables. First, it reveals that the Algorithm, with 20.83\%, is the most influential factor on the Latency SLA Violation response variable. However, there is a notable balance between all factors, with the combination of Workload and Cloud at 14.86\%.

Figure \ref{fig:influence_combined} also shows that the Cloud factor, with 43.40\%, has the major influence on the Average Latency response variable. The Algorithm, at 24.62\%, is also significant. The Workload factor appears less relevant in this scenario, likely due to the trade-off between latency and resources in edge-cloud infrastructure.


Regarding Power Consumption, Figure \ref{fig:influence_combined} indicates that the Cloud factor, at 32.52\%, is the most influential. However, the combination of Algorithm and Workload, at 40.10\%, suggests these factors are also relevant. This behavior is attributed to the relevance of cloud presence in terms of energy consumption.

\begin{figure}[ht!]
    \centering
    \scriptsize
    \begin{tikzpicture}
        \begin{axis}[
            ybar,
            clip=false, 
            symbolic x coords={
                Exp1\\Tetris\\Low\\On,
                Exp2\\Tetris\\Low\\Off,
                Exp3\\Tetris\\High\\On,
                Exp4\\Tetris\\High\\Off,
                Exp5\\Thea\\Low\\On,
                Exp6\\Thea\\Low\\Off,
                Exp7\\Thea\\High\\On,
                Exp8\\Thea\\High\\Off
            },
            xtick=data,
            ymin=0,
            ymax=300,
            ylabel={Latency SLA Violations},
            bar width=15pt,
            enlarge x limits=0.07,
            nodes near coords,
            every node near coord/.append style={font=\scriptsize, yshift=2pt, text=black},
            xticklabel style={align=center},
            width=8cm,
            height=4.5cm
        ]
        \addplot+[fill=blue!50, draw=black, error bars/.cd, y dir=both, y explicit, error bar style={black}] coordinates {
            (Exp1\\Tetris\\Low\\On,  18.70) +- (4,4)
            (Exp2\\Tetris\\Low\\Off, 18.60) +- (4.05,4.05)
            (Exp3\\Tetris\\High\\On, 22.20) +- (3.4,3.4)
            (Exp4\\Tetris\\High\\Off,24.70) +- (3,3)
            (Exp5\\Thea\\Low\\On,    40.60) +- (5.15,5.15)
            (Exp6\\Thea\\Low\\Off,   35.10) +- (3.55,3.55)
            (Exp7\\Thea\\High\\On,   29.60) +- (1.2,1.2)
            (Exp8\\Thea\\High\\Off,  253.0) +- (22.25,22.25)
        };
        
        \draw
          (axis description cs:-0.21,-0.45)
          node[anchor=south west, align=left] {\scriptsize
            \textbf{Experiment:}\\
            \textbf{Algorithm:}\\
            \textbf{Workload:}\\
            \textbf{Cloud:}
          };
        \end{axis}
    \end{tikzpicture}
    \caption{Latency SLA Violations (Deadlines) performance by experiment}
    \label{fig:exp_by_sla}
\end{figure}
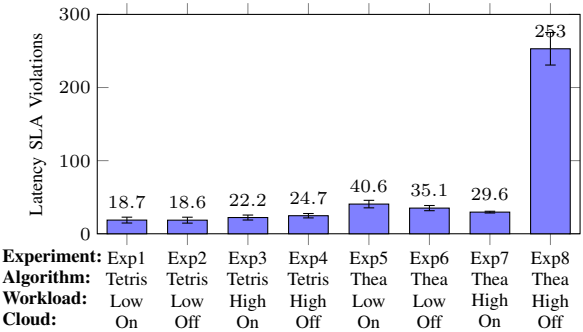
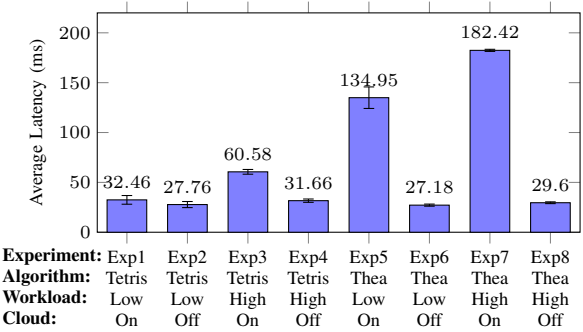
\begin{figure}[ht!]
    \centering
    \scriptsize
    \begin{tikzpicture}
        \begin{axis}[
            ybar,
            clip=false, 
            symbolic x coords={
                Exp1\\Tetris\\Low\\On,
                Exp2\\Tetris\\Low\\Off,
                Exp3\\Tetris\\High\\On,
                Exp4\\Tetris\\High\\Off,
                Exp5\\Thea\\Low\\On,
                Exp6\\Thea\\Low\\Off,
                Exp7\\Thea\\High\\On,
                Exp8\\Thea\\High\\Off
            },
            xtick=data,
            ymin=0,
            ymax=220,
            ylabel={Average Latency (ms)},
            bar width=15pt,
            enlarge x limits=0.07,
            nodes near coords,
            every node near coord/.append style={font=\scriptsize, yshift=2pt, text=black},
            xticklabel style={align=center},
            width=8cm,
            height=4.5cm
        ]
        \addplot+[fill=blue!50, draw=black, error bars/.cd, y dir=both, y explicit, error bar style={black}] coordinates {
            (Exp1\\Tetris\\Low\\On,  32.46) +- (4.36,4.36)
            (Exp2\\Tetris\\Low\\Off, 27.76) +- (3.05,3.05)
            (Exp3\\Tetris\\High\\On, 60.58) +- (2.38,2.38)
            (Exp4\\Tetris\\High\\Off,31.66) +- (1.67,1.67)
            (Exp5\\Thea\\Low\\On,    134.95) +- (10.8,10.8)
            (Exp6\\Thea\\Low\\Off,   27.18) +- (1.06,1.06)
            (Exp7\\Thea\\High\\On,   182.42) +- (1.06,1.06)
            (Exp8\\Thea\\High\\Off,  29.6) +- (0.9,0.9)
        };
        \draw
          (axis description cs:-0.21,-0.45)
          node[anchor=south west, align=left] {\scriptsize
            \textbf{Experiment:}\\
            \textbf{Algorithm:}\\
            \textbf{Workload:}\\
            \textbf{Cloud:}
          };
        \end{axis}
    \end{tikzpicture}
    \caption{Average Latency performance by experiment}
    \label{fig:exp_by_avg}
\end{figure}
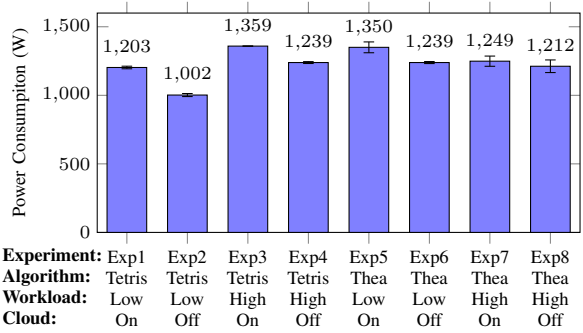
\begin{figure}[ht!]
    \centering
    \scriptsize
    \begin{tikzpicture}
        \begin{axis}[
            ybar,
            clip=false, 
            symbolic x coords={
                Exp1\\Tetris\\Low\\On,
                Exp2\\Tetris\\Low\\Off,
                Exp3\\Tetris\\High\\On,
                Exp4\\Tetris\\High\\Off,
                Exp5\\Thea\\Low\\On,
                Exp6\\Thea\\Low\\Off,
                Exp7\\Thea\\High\\On,
                Exp8\\Thea\\High\\Off
            },
            xtick=data,
            ymin=0,
            ymax=1600,
            ylabel={Power Consumpiton (W)},
            bar width=15pt,
            enlarge x limits=0.07,
            nodes near coords,
            every node near coord/.append style={font=\scriptsize, yshift=2pt, text=black},
            xticklabel style={align=center},
            width=8cm,
            height=4.5cm
        ]
        \addplot+[fill=blue!50, draw=black, error bars/.cd, y dir=both, y explicit, error bar style={black}] coordinates {
            (Exp1\\Tetris\\Low\\On,  1203) +- (9.41,9.41)
            (Exp2\\Tetris\\Low\\Off, 1002) +- (11.04,11.04)
            (Exp3\\Tetris\\High\\On, 1359 ) +- (0.15,0.15)
            (Exp4\\Tetris\\High\\Off,1239) +- (6.92,6.92)
            (Exp5\\Thea\\Low\\On,    1350 ) +- (39.52,39.52)
            (Exp6\\Thea\\Low\\Off,   1239) +- (7.4,7.4)
            (Exp7\\Thea\\High\\On,   1249) +- (37.24,37.24)
            (Exp8\\Thea\\High\\Off,  1212) +- (46.09,46.09)
        };
        \draw
          (axis description cs:-0.21,-0.45)
          node[anchor=south west, align=left] {\scriptsize
            \textbf{Experiment:}\\
            \textbf{Algorithm:}\\
            \textbf{Workload:}\\
            \textbf{Cloud:}
          };
        \end{axis}
    \end{tikzpicture}
    \caption{Power Consumption performance by experiment}
    \label{fig:exp_by_power}
\end{figure}

Another relevant element from the experiment planning results is the behavior of response variables across different experiment setups. These are shown in Figures \ref{fig:exp_by_sla}, \ref{fig:exp_by_avg}, and \ref{fig:exp_by_power}.

Figure \ref{fig:exp_by_sla} indicates that the last experiment setup had the worst performance regarding Latency SLA Violations. Specifically, Experiment 8 recorded approximately 250 violations, significantly higher than the fewer than 50 violations reported in the first seven experiments. This highlights the differing performances of the algorithms under various conditions. Thea exhibits significant performance deterioration under high workload conditions and without the cloud option, leading to a substantial increase in Latency SLA Violations. Conversely, Tetris maintains consistent performance across different workloads and infrastructure scenarios.

Figure \ref{fig:exp_by_avg} allows for a critical discussion on edge-cloud infrastructure. Notably, the first four setups (Tetris experiments) show that experiments 1 and 3 have higher Average Latency values. For the last four setups (Thea experiments), experiments 5 and 7 exhibit higher Average Latency values than all eight setups, indicating the influence of Edge-Cloud infrastructure. As mentioned earlier, the trade-off between resource availability and latency explains this behavior. Although the presence of the cloud increases average latency, it does not necessarily imply a negative impact on SLA violations.

In terms of Power Consumption, Figure \ref{fig:exp_by_power} compares all setups and shows that, while there is no significant difference in average values between Tetris and Thea, Thea exhibits higher variation, particularly in experiments 5, 7, and 8. Tetris, on the other hand, maintains more consistent power consumption values.

Furthermore, Table \ref{tab:algo_quantitative} shows the behavior of response variables for the Tetris and Thea algorithms. These values result from the mean and standard deviation calculations of all executed experiments. Tetris performs significantly better in terms of latency SLA violations and drop SLA violations than Thea. Specifically, Thea records around 90 latency violations, whereas Tetris has fewer than 30. In terms of drops, Tetris has no occurrences, whereas Thea does.

Additionally, Table \ref{tab:algo_quantitative} highlights the dispersion of each measure. For both Latency and Drop violations, the dispersion of Thea values is much larger than that of Tetris. This observation reinforces the findings from Figures \ref{fig:exp_by_sla}, \ref{fig:exp_by_avg}, and \ref{fig:exp_by_power}, where Thea's performance varied significantly during the experiments, while Tetris maintained consistent performance. Table \ref{tab:algo_quantitative} also shows that Thea's Average Latency is twice that of Tetris, indicating that Thea accesses the cloud server more frequently. Regarding Power Consumption, there is no significant statistical difference between Thea and Tetris, with both averaging around 1200 and 1300 Watts.

\begin{table}[ht!]
\footnotesize
  \setlength{\tabcolsep}{2.20pt}
  \renewcommand{\arraystretch}{1.15}
  \caption{Algorithms performance for each response variable under all executed experiments.}
  \label{tab:algo_quantitative}
  \begin{center}
  \begin{tabular}{cc|c|c|c}
    \cline{1-5}
    \multirow{2}{*}{Metrics} & \multicolumn{2}{c|}{Tetris} & \multicolumn{2}{c}{Thea} \\ 
    \cline{2-5}
              & mean & std & mean & std \\ 
    \cline{1-5}
    Latency SLA Violations & 21.5 & 6.5 & 89.6 & 97.6 \\ 
    Drop SLA Violations & 0 & 0 & 0.4 & 0.7 \\ 
    Average Latency (ms) & 38.1 & 14.2 & 93.5 & 68.8 \\ 
    Power Consumption (W) & 1201.4 & 131.1 & 1263.0 & 77.8 \\ 
    \cline{1-5}
  \end{tabular}
  \end{center}
\end{table}

Therefore, Table \ref{tab:algo_qualitative} presents a qualitative comparison. Except for parity regarding Power Consumption, the table highlights Tetris's superior performance, particularly for SLA Violations variables.

\begin{table}[ht!]
\footnotesize
  \centering
  \caption{Qualitative Comparison: Tetris x Thea.}
  \label{tab:algo_qualitative}
  \begin{tabular}{ccc}
    \toprule
    Variable&Tetris&Thea\\
    \midrule
    Latency SLA Vio. & Much Better &  Much Worse   \\ 
    Drop SLA Vio. & Much Better &  Much Worse   \\ 
    Avg. Latency & Better & Worse  \\ 
    Power Consumption & Parity & Parity  \\
  \bottomrule
\end{tabular}
\end{table}

Another important comparison is between the Edge-Cloud and Edge-Only architectures. Table \ref{tab:cloud_quantitative} provides relevant insights into this investigation. In terms of violation response variables (Latency and Drop), cloud availability is crucial for mitigating violations. When the cloud is off, there is a high mean value (around 80 Latency SLA violations and some Drop occurrences) with significant dispersion (exceeding 100 Latency SLA Violations and 0.5 Drop SLA Violations). When the cloud is on, the mean Latency SLA Violations are reduced by half, the dispersion is much smaller, and no Drop SLA Violations occur.

In addition, since sending applications to be processed on the cloud introduces more latency than sending to edge servers, the Average Latency response variable behaves as expected, as shown in Table \ref{tab:cloud_quantitative}. This increased latency does not necessarily imply negative effects on agreement violations. Power Consumption is slightly higher when the cloud is active, which is understandable since cloud servers consume more energy than edge servers.

\begin{table}[ht!]
\footnotesize
  \setlength{\tabcolsep}{2.20pt}
  \renewcommand{\arraystretch}{1.25}
  \caption{Architecture performance for each response variable under all executed experiments.}
  \label{tab:cloud_quantitative}
  \begin{center}
  \begin{tabular}{cc|c|c|c}
    \cline{1-5}
    \multirow{2}{*}{Metrics} & \multicolumn{2}{c|}{Edge-Cloud} & \multicolumn{2}{c}{Edge-Only} \\ 
    \cline{2-5}
              & mean & std & mean & std \\ 
    \cline{1-5}
    Latency SLA Violations & 27.8 & 10.6 & 82.8 & 101.6 \\ 
    Drop SLA Violations & 0 & 0 & 0.4 & 0.7 \\ 
    Average Latency (ms) & 102.6 & 61.0 & 29.0 & 3.5 \\ 
    Power Consumption (W) & 1290.9 & 80.3 & 1173.5 & 108.0 \\ 
    \cline{1-5}
  \end{tabular}
  \end{center}
\end{table}

Table \ref{tab:cloud_qualitative} compares the Edge-Cloud and Edge-Only infrastructures, summarizing the response variables' behavior for both architectures. For SLA Violations variables, Edge-Cloud demonstrates superior performance. However, in terms of Average Latency, Edge-Only has the advantage. Regarding Power Consumption, the presence of the cloud results in a slight increase.

\begin{table}[ht!]
\small
  \centering
  \caption{Qualitative Comparison: Edge-Cloud x Edge-Only.}
  \label{tab:cloud_qualitative}
  \begin{tabular}{ccc}
    \toprule
    Variable&Edge-Cloud&Edge-Only\\
    \midrule
    Latency SLA Vio. & Much Better &  Much Worse   \\ 
    Drop SLA Vio. & Much Better &  Much Worse   \\ 
    Avg. Latency & Much Worse & Much Better  \\ 
    Power Consumption & Slightly Worse & Slightly Better  \\
  \bottomrule
\end{tabular}
\end{table}

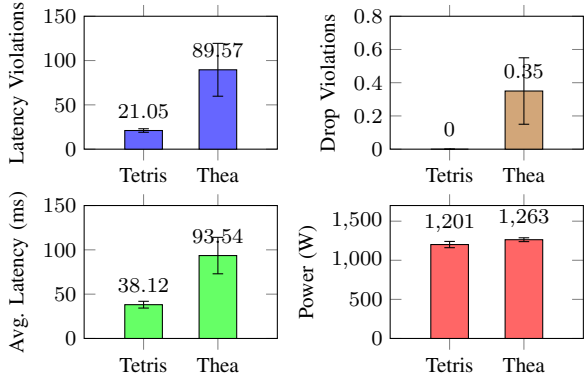
\begin{figure}[ht!]
    \centering
    \footnotesize
    \begin{tikzpicture}
    \begin{groupplot}[
        group style={
            group size=2 by 2,     
            horizontal sep=1.5cm,  
            vertical sep=0.75cm   
        },
        width=4.15cm,  
        height=3.35cm, 
        ybar,                    
        symbolic x coords={Tetris,Thea}, 
        xtick=data,
        enlarge x limits=0.8,      
        nodes near coords,
        every node near coord/.append style={font=\footnotesize, yshift=2pt, text=black}
    ]

    \nextgroupplot[
        ylabel={Latency Violations},
        ymin=0, ymax=150,
        bar width=5mm  
    ]
    \addplot+[fill=blue!60, draw=black, error bars/.cd, y dir=both, y explicit, error bar style={black}] coordinates {
        (Tetris, 21.05) +- (2.01,2.01)
        (Thea,   89.57) +- (29.82,29.82)
    };

    \nextgroupplot[
        ylabel={Drop Violations},
        ymin=0, ymax=0.80,
        bar width=5mm
    ]
    \addplot+[fill=brown!70, draw=black, error bars/.cd, y dir=both, y explicit, error bar style={black}] coordinates {
        (Tetris, 0) +- (0,0)
        (Thea,   0.35) +- (0.2,0.2)
    };

    \nextgroupplot[
        ylabel={Avg. Latency (ms)},
        ymin=0, ymax=150,
        bar width=5mm
    ]
    \addplot+[fill=green!60, draw=black, error bars/.cd, y dir=both, y explicit, error bar style={black}] coordinates {
        (Tetris, 38.12)  +- (3.82,3.82)
        (Thea,   93.54) +- (20.57,20.57)
    };

    \nextgroupplot[
        ylabel={Power (W)},
        ymin=0, ymax=1700,
        bar width=5mm
    ]
    \addplot+[fill=red!60, draw=black, error bars/.cd, y dir=both, y explicit, error bar style={black}] coordinates {
        (Tetris, 1201) +- (40.1, 40.1)
        (Thea,   1263) +- (24.58, 24.58)
    };

    \end{groupplot}
    \end{tikzpicture}
    \caption{Comparison of Tetris and Thea across all response variables under every experimental configuration.}
    \label{fig:algo_comparison}
\end{figure}

Figure~\ref{fig:algo_comparison} presents the experimental results for the response variables, such as Latency SLA Violations, Drop SLA Violations, Average Latency, and Power Consumption, across all experimental configurations. The data are organized to facilitate a direct comparison between Tetris and the state-of-the-art algorithm, Thea, providing a quantitative baseline for performance evaluation. This analysis shows that Tetris has lower Latency SLA Violations and no Drop SLA Violations, while Thea has higher values for both metrics. These results quantify a reduction of approximately 76.5\% in latency violations when using Tetris, and they confirm that Tetris achieves more consistent compliance with SLA requirements under the tested conditions.


\begin{figure}[ht!]
    \centering
    \footnotesize
    \begin{tikzpicture}
    \begin{groupplot}[
        group style={
            group size=2 by 2,     
            horizontal sep=1.5cm,   
            vertical sep=0.75cm     
        },
        width=4.15cm,  
        height=3.35cm,   
        ybar,                     
        symbolic x coords={Low,High}, 
        xtick=data,
        enlarge x limits=0.8,      
        nodes near coords,
        every node near coord/.append style={font=\footnotesize, yshift=2pt, text=black}
    ]

    \nextgroupplot[
        ylabel={Latency Violations},
        ymin=0, ymax=150,
        bar width=5mm  
    ]
    \addplot+[fill=blue!60, draw=black, error bars/.cd, y dir=both, y explicit, error bar style={black}] coordinates {
        (Low,  28.25) +- (3.5,3.5)
        (High, 82.375) +- (29.9,29.9)
    };

    \nextgroupplot[
        ylabel={Drop Violations},
        ymin=0, ymax=0.80,
        bar width=5mm
    ]
    \addplot+[fill=brown!70, draw=black, error bars/.cd, y dir=both, y explicit, error bar style={black}] coordinates {
        (Low, 0) +- (0,0)
        (High,   0.35) +- (0.2,0.2)
    };

    \nextgroupplot[
        ylabel={Avg. Latency (ms)},
        ymin=0, ymax=120,
        bar width=5mm
    ]
    \addplot+[fill=green!60, draw=black, error bars/.cd, y dir=both, y explicit, error bar style={black}] coordinates {
        (Low, 55.59)  +- (15.36,15.36)
        (High, 76.06) +- (18.75,18.75)
    };

    \nextgroupplot[
        ylabel={Power (W)},
        ymin=0, ymax=1700,
        bar width=5mm
    ]
    \addplot+[fill=red!60, draw=black, error bars/.cd, y dir=both, y explicit, error bar style={black}] coordinates {
        (Low, 1199) +- (39.9, 39.9)
        (High, 1265) +- (22.08, 22.08)
    };

    \end{groupplot}
    \end{tikzpicture}
    \caption{Effect of workload level (Low vs. High) on the response variables across all executed experiments.}
    \label{fig:workload_comparison}
\end{figure}
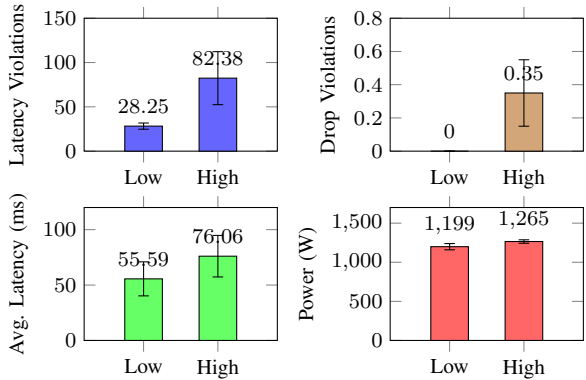

Figure~\ref{fig:workload_comparison} illustrates the performance results corresponding to varying workload conditions. The figure compares system performance under Low and High workload levels. As workload increases, Latency Violations rise from approximately 28 to 90, and Drop Violations increase from 0 to about 0.35, indicating a higher rate of missed deadlines and dropped requests. Average Latency also grows from about 56 ms to 76 ms, reflecting longer completion times under heavier demand. Finally, Power Consumption increases from roughly 1198 W to 1265 W, showing that higher workloads lead to greater energy usage.


\begin{figure}[ht!]
    \centering
    \footnotesize
    \begin{tikzpicture}
    \begin{groupplot}[
        group style={
            group size=2 by 2,     
            horizontal sep=1.5cm,   
            vertical sep=0.75cm      
        },
        width=4.15cm,  
        height=3.35cm,  
        ybar,                      
        symbolic x coords={On,Off}, 
        xtick=data,
        enlarge x limits=0.8,      
        nodes near coords,
        every node near coord/.append style={font=\footnotesize, yshift=2pt, text=black}
    ]

    \nextgroupplot[
        ylabel={Latency Violations},
        ymin=0, ymax=150,
        bar width=5mm  
    ]
    \addplot+[fill=blue!60, draw=black, error bars/.cd, y dir=both, y explicit, error bar style={black}] coordinates {
        (On, 27.77) +- (3.01,3.01)
        (Off, 82.85) +- (30.9,30.9)
    };

    \nextgroupplot[
        ylabel={Drop Violations},
        ymin=0, ymax=0.80,
        bar width=5mm
    ]
    \addplot+[fill=brown!70, draw=black, error bars/.cd, y dir=both, y explicit, error bar style={black}] coordinates {
        (On, 0) +- (0,0)
        (Off,   0.35) +- (0.2,0.2)
    };

    \nextgroupplot[
        ylabel={Avg. Latency (ms)},
        ymin=0, ymax=160,
        bar width=5mm
    ]
    \addplot+[fill=green!60, draw=black, error bars/.cd, y dir=both, y explicit, error bar style={black}] coordinates {
        (On, 102.61)  +- (18.14,18.14)
        (Off, 29.05) +- (1.05,1.05)
    };

    \nextgroupplot[
        ylabel={Power (W)},
        ymin=0, ymax=1700,
        bar width=5mm
    ]
    \addplot+[fill=red!60, draw=black, error bars/.cd, y dir=both, y explicit, error bar style={black}] coordinates {
        (On, 1291) +- (24.83, 24.83)
        (Off, 1173) +- (33.66, 33.66)
    };

    \end{groupplot}
    \end{tikzpicture}
    \caption{Impact of cloud availability (On vs. Off) on the response variables across all executed experiments.}
    \label{fig:cloud_comparison}
\end{figure}
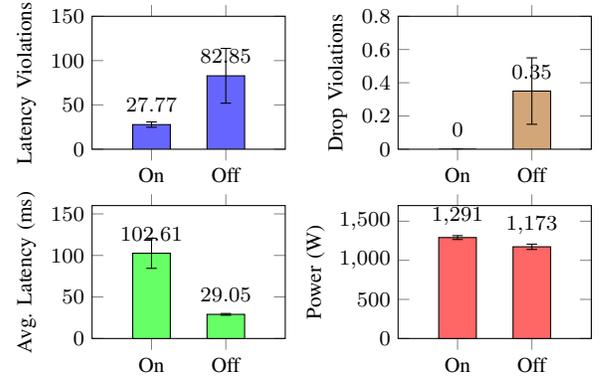

Figure~\ref{fig:cloud_comparison} displays the system performance metrics under different cloud availability scenarios. The figure segregates results for cloud-on and cloud-off configurations, thereby highlighting the impact of cloud resources on the measured response variables. When the cloud is On, Latency Violations drop from about 83 to 28, and there are no Drop Violations, indicating more reliable SLA compliance. However, Average Latency rises from roughly 29 ms to 103 ms, reflecting the added distance to cloud resources. Additionally, Power Consumption increases from about 1173 W to 1290 W, suggesting that activating the cloud server requires more energy overall.


Table \ref{tab:final_results} presents the mean and standard deviation values for each response variable across all experiment setups. Concerning the algorithms, Tetris is, on average, 76.50\% better than Thea for the Latency SLA Violations response variable. Considering the standard deviation, Tetris's performance ranges between 85.70\% and 63.84\% better. Regarding Drop SLA Violations, the data similarly indicate Tetris's superiority; it registers zero instances of drops, whereas Thea experiences an average of 1.4 application drops in experiment 8. 

Focusing on cloud availability, it is interesting to note that, in terms of SLA Violations variables, Edge-Cloud, aside from avoiding drop occurrences, is also 66.47\% superior to Edge-Only in Latency SLA Violations, with an improvement interval ranging from 77.65\% in the best case to 50.61\% in the worst case.

\begin{table}[ht!]
\setlength{\tabcolsep}{3pt} 
\renewcommand{\arraystretch}{1.3} 
\caption{Performance metrics (latency violations, drops, average latency, and power) for Tetris and Thea under different load (low/high) and cloud (on/off) conditions.}
\label{tab:final_results}
\begin{center}
\adjustbox{width=0.99\linewidth}{
\begin{tabular}{cccccc|cc|cc|cc}
\hline
\multirow{2}{*}{Exp} & \multirow{2}{*}{Algo} & \multirow{2}{*}{Load} & \multirow{2}{*}{Cloud} & \multicolumn{2}{c|}{Latency Vio.} & \multicolumn{2}{c|}{Drops} & \multicolumn{2}{c|}{Avg. Latency} & \multicolumn{2}{c}{Power} \\ \cline{5-12} 
                              &                                &                                &                                 & \multicolumn{1}{c|}{mean}     & std      & \multicolumn{1}{c|}{mean}   & std   & \multicolumn{1}{c|}{mean}      & std      & \multicolumn{1}{c|}{mean}     & std \\ \hline
1                             & Tetris                         & low                            & on                              & \multicolumn{1}{c|}{18.7}     & 6.9     & \multicolumn{1}{c|}{0.0}    & 0.0   & \multicolumn{1}{c|}{32.5}     & 7.5     & \multicolumn{1}{c|}{1203.8}  & 16.8  \\ 
2                             & Tetris                         & low                            & off                             & \multicolumn{1}{c|}{18.6}     & 6.9     & \multicolumn{1}{c|}{0.0}    & 0.0   & \multicolumn{1}{c|}{27.8}     & 5.3     & \multicolumn{1}{c|}{1002.1}  & 19.8 \\ 
3                             & Tetris                         & high                           & on                              & \multicolumn{1}{c|}{22.2}     & 5.9     & \multicolumn{1}{c|}{0.0}    & 0.0   & \multicolumn{1}{c|}{60.6}     & 3.9     & \multicolumn{1}{c|}{1359.8}  & 0.3 \\ 
4                             & Tetris                         & high                           & off                             & \multicolumn{1}{c|}{24.7}     & 5.1     & \multicolumn{1}{c|}{0.0}    & 0.0   & \multicolumn{1}{c|}{31.7}     & 2.7     & \multicolumn{1}{c|}{1239.9}  & 11.9 \\ \hline
5                             & Thea                           & low                            & on                              & \multicolumn{1}{c|}{40.6}     & 9.1     & \multicolumn{1}{c|}{0.0}    & 0.0   & \multicolumn{1}{c|}{135.0}    & 19.9    & \multicolumn{1}{c|}{1350.1}  & 66.0 \\ 
6                             & Thea                           & low                            & off                             & \multicolumn{1}{c|}{35.1}     & 6.2     & \multicolumn{1}{c|}{0.0}    & 0.0   & \multicolumn{1}{c|}{27.2}     & 1.8     & \multicolumn{1}{c|}{1239.8}  & 12.5 \\ 
7                             & Thea                           & high                           & on                              & \multicolumn{1}{c|}{29.6}     & 2.3     & \multicolumn{1}{c|}{0.0}    & 0.0   & \multicolumn{1}{c|}{182.4}    & 1.9     & \multicolumn{1}{c|}{1250.0}  & 62.1 \\ 
8                             & Thea                           & high                           & off                             & \multicolumn{1}{c|}{253.0}    & 39.2    & \multicolumn{1}{c|}{1.4}    & 0.5  & \multicolumn{1}{c|}{29.6}      & 1.6     & \multicolumn{1}{c|}{1212.3}  & 76.1 \\ \hline
\end{tabular}
}
\end{center}
\end{table}

Even though Tetris is not significantly superior to the state-of-the-art approach in terms of the power consumption response variable, simulated results have demonstrated that Tetris achieves near-optimal results for Service-Level Agreement variables. Under varying workload demands and different cloud availability conditions, Tetris reduces Latency SLA Violations by an average of 76.50\% and completely avoids Drop SLA Violations.


\section{Conclusion}
\label{sec:conclusion}

This paper presented Tetris, an SLA-aware application placement strategy for the edge-cloud continuum. Unlike existing approaches, such as the state-of-the-art Thea, which optimizes resource allocation through proximity heuristics or static placement rules, Tetris adapts to workload fluctuations. It continuously optimizes resource distribution to maintain efficiency and prevent performance degradation. Tasks are prioritized according to their SLA requirements, reducing latency, preventing application drops, and maintaining energy efficiency without compromising deadline compliance. 

Experimental results indicate that Tetris achieves a 76.50\% reduction in latency violations and eliminates application drops in all evaluated scenarios. Additionally, Tetris ensures deadline compliance while maintaining power consumption levels comparable to Thea. By integrating latency, deadline, drop rate, and energy consumption as optimization criteria, Tetris provides a systematic SLA-aware placement strategy, effectively managing workload variability and resource constraints in edge-cloud infrastructures.


\bibliography{referencias}

\bibliographystyle{IEEEtran} 

\end{document}